\documentstyle[12pt,aasms4]{article}

\newcommand{\etal}{{\it et al.}}
\newcommand{\eg}{{\it e.g.}}
\newcommand{\ie}{{\it i.e.}}

\newcommand{\msun}{${\rm M}_\odot$}
\newcommand{\ps}{s$^{-1}$}
\newcommand{\mspyr}{${\rm M}_\odot$ yr$^{-1}$}
\newcommand{\scinot}[2]{${#1}{\times}10^{{#2}}$}
\newcommand{\tento}[1]{$10^{{#1}}$}
\newcommand{\telone}{1.3$^{\rm m}$\ }
\newcommand{\teltwo}{2.4$^{\rm m}$\ }
\newcommand{\telct}{1.5$^{\rm m}$\  }
\def\la{\mathrel{\hbox{\rlap{\hbox{\lower4pt\hbox{$\sim$}}}\hbox{$<$}}}}
\def\ga{\mathrel{\hbox{\rlap{\hbox{\lower4pt\hbox{$\sim$}}}\hbox{$>$}}}}

\received{revised version, ApJ, 1998 May 5}
\begin{document}

\title{Bipolar Jets and Orbital Dynamics of the Supersoft X-Ray Source RX J0019.8+2156} 
\author{C. M. Becker, R. A. Remillard, S. A. Rappaport,} 
\affil{Center for Space Research, Massachusetts
Institute of Technology, Cambridge, MA 02139}
\authoremail{cmbecker@space.mit.edu, rr@space.mit.edu,
sar@space.mit.edu} 
\author{J. E. McClintock}
\affil{Harvard-Smithsonian Center for Astrophysics, 60 Garden St.,
Cambridge, MA 02138} 
\authoremail{jem@head-cfa.harvard.edu}

\begin{abstract}
Between 1994 July and 1997 February we monitored the optical spectrum
of RX J0019.8+2156. This supersoft X-ray source is one of only two
accreting white dwarfs in the Galaxy that are thought to be burning
hydrogen on their surface as a consequence of a high rate of mass
transfer from a binary companion. Accurate orbital ephemerides are
derived from radial velocity measurements for the white dwarf, which
are obtained from strong He\,II emission lines with a stable velocity
semi-amplitude of $K = 71.2 \pm 3.6$ km \ps. We report the discovery
of transient, low-velocity, bipolar jets. These jets are respresented
by redshifted-blueshifted pairs of emission lines from H and He\,II
with an outflow velocity of $v$ cos($i$) $\sim$ 815 km \ps, where $i$
is the binary inclination angle. When present, the jet lines seen in
$H\alpha$ also exhibit an orbital modulation of 71 km \ps\, which
strengthens the interpretation that this is the orbital velocity of
the white dwarf and also indicates that the jets are oriented nearly
perpendicular to the orbital plane. On most occasions, the H emission
line profiles are further altered by P Cygni absorption effects, and
the strength of this absorption is also dependent on binary phase.  We
show that the jets and the P Cygni features have very different
temporal characteristics and binary phase dependence; thus, we conclude
that the outflowing material and the absorbing wind must have
essentially different geometries. Finally, the measured mass function
is combined with binary evolution models to suggest a limit on the
inclination angle, $i < 40^\circ$.  A particular model invoked to
explain a high rate of mass transfer requires $16^\circ < i <
25^\circ$. However, at such small inclination it is difficult to
explain the large amplitude of the orbital light curve ($\sim 0.5$
mag). Alternatively, the results could signify substantial vertical
structure in the accretions disks of supersoft X-ray sources. By
contrast, a simple model fit to the jet outflow lines indicates an
orbital inclination angle of $35^\circ < i < 60^\circ$.
\end{abstract}


\section{Introduction}

Supersoft X-ray sources exhibit extremely steep spectra in the range
of 0.1 - 2 keV. They can be divided into three general subclasses:
low and high luminosity galactic objects, and AGN. In this paper we
concern ourselves with luminous galactic supersoft X-ray sources
(hereafter SXS), which have characteristic luminosities of
$\sim$\tento{36} - \tento{38} erg \ps\ and effective temperatures of
$\sim$2-\scinot{6}{5} K ($kT \simeq 17-50$ eV).  The first luminous
supersoft sources were discovered during a survey of the Large
Magellanic Cloud with the Einstein satellite (\cite{cal83:disc}).  The
ROSAT X-ray survey (\cite{rass:summary}) has now established that the
SXS constitute a distinct astronomical class.  Greiner (1996)
\nocite{sxscat:greiner} has cataloged and summarized the properties of
34 known SXS.  Only seven of these objects are in the Galaxy, while
seven are in the LMC, four are in the SMC, and sixteen have been
located in M31.

 A particular model for at least some of the long-lived SXS invokes
steady or cyclical nuclear burning of matter on the surface of a
$\sim$1 \msun\ white dwarf that is accreting mass from a main-sequence
or subgiant companion star of $\sim 1 - 2$ \msun\ at a rate of
\scinot{3}{-8} $-$ \tento{-6} \mspyr\ (\cite{nbwd:hbnr92}, hereafter
HBNR; \cite{sxs:popsynth:rds94}, hereafter RDS; and references
therein).  Such high transfer rates are a natural consequence of
thermal-timescale mass transfer via Roche lobe overflow, which
characteristically occurs in this type of system.  The transfer is
potentially unstable because the donor star is more massive than the
accreting white dwarf; however, the rate of transfer is limited by the
thermal time scale of the donor star.

Models for the evolution of these systems (RDS) indicate that the
orbital periods should lie in the range of 8 hr to 1.4 d.  Moreover,
these studies predict that there should be some $10^3$ such systems in
the Galaxy and in M31.  In spite of the greater distance, it is
apparently easier to detect these supersoft sources (at least in the
soft X-ray band) in nearby external galaxies at relatively high
$b^{II}$, such as the LMC ($-33^\circ$) and M31 ($-21^\circ$).  In the
plane of our own Galaxy, a hydrogen column density of only
${\sim}10^{21}$ cm$^{-2}$ will absorb most of the very soft
X-radiation.  In the Galactic plane this corresponds to distances of
less than a kpc.

Five of the seven SXS in the Milky Way are {\it un}likely to be the type of
nuclear burning white dwarf (NBWD) discussed above.  Two are recently
exploded novae, two are symbiotic novae, and one is unclassified. Thus,
there are only two good NBWD candidates in the Galaxy: RX~J0925.7-4758
(\cite{0925:disc}) and RX~J0019.8+2156 (hereafter RXJ0019;
\cite{0019:disc}).  Since RX~J0925.7-4758 is obscured by the Vela
Sheet molecular cloud, and its intrinsic optical spectrum is heavily
reddened (\cite{0925:disc}; \cite{0925:absorber}), RXJ0019 remains as
the one prime galactic NBWD candidate for detailed optical study.

Beuermann \etal\ (1995) \nocite{0019:disc} discovered RXJ0019 as part
of the ROSAT all-sky survey in 1990 December. They subsequently
initiated a battery of short ROSAT PSPC, ROSAT HRI, IUE, and optical
observations in 1992-93 (\cite{0019:disc}). The following is a brief
summary of their results. The optical counterpart to RXJ0019 is a
moderately bright object with $V=12.2$. The light curves of the X-ray,
optical, and UV data show modulations with a period of 15.85 hr,
interpreted as the binary period, while smaller, aperiodic
fluctuations are seen on time scales of hours. They adopt a distance
to RXJ0019 of 2 kpc on the basis of the calcium absorption features
and the X-ray column density ($N_H \sim$ \scinot{4.4}{20}
cm$^{-2}$). The X-ray spectrum is extremely soft and luminous. For an
assumed LTE model atmosphere (\cite{nbwd:lte-model}), the ROSAT
spectrum yields $kT = 34$ eV and an estimated bolometric luminosity,
$L_{bol} \simeq$ \scinot{4}{36} erg \ps\, with significant uncertainty
caused by the limited spectral constraints.  The optical spectrum
shows strong hydrogen lines, exceptionally strong He\,II emission, and
P Cygni absorption associated with the Balmer lines. Beuermann \etal\
also report that the radial velocity of the He\,II $\lambda$4686{\AA}
line is modulated at the orbital period with a semi-amplitude of $67
\pm 3$ km \ps.
 
Greiner \& Wenzel (1995) \nocite{0019:hps} examined archival
photographic plates dating between 1890 and 1990 from the Harvard
plate stacks and the Sonnenberg Observatory. They detected the orbital
period, low amplitude variations on time scales as short as 2 hr, and
$\sim$1 mag variations over intervals of $\sim$20 years.

Will \& Barwig (1996) \nocite{0019:photometry} studied RXJ0019 with
UBVRI photometry from 1992 to 1995 using a multichannel photometer on
the Wendelstein Observatory 80-cm telescope.  RXJ0019 shows both
primary and secondary minima; the primary minimum is a $\sim 0.5^{\rm
m}$ eclipse which covers nearly 40\% of the orbital phase (FWHM).
Will \& Barwig's analysis of mid-eclipse times provides an accurate
orbital ephemeris with a period of $15.85133 \pm 0.00017$ hr.

G\"{a}nsicke, Beuermann, \& de\,Martino (1996) \nocite{0019:uvspec}
observed RXJ0019 with the IUE telescope using both the
short-wavelength prime (SWP; 1250-1900 \AA) and long wavelength prime
(LWP; 2000-3200 \AA) spectrographs.  The 15.85 hr periodicity, and
both primary and secondary dips are present in the UV light curves;
primary minima were deeper in the LWP spectra (32\% modulation) than
in the SWP spectra (22\%).  The He\,II $1640$\AA\ line is present and
shows a stronger modulation than the continuum (43\%).  Fits to the UV
spectrum using a simple model for an irradiated donor and accretion
disk (\cite{0513:irrad}) yield parameters consistent with masses of
$\sim$1.0 and $\sim$1.5 \msun\ for the white dwarf and donor star,
respectively (\cite{0019:uvspec}).

In this work we present our spectroscopic observations of RXJ0019.
The radial velocity work of Beuermann \etal\ (1995),
\nocite{0019:disc} was completed in four days. In contrast, our
observations span $\sim$ 2.5 yr and allow us to probe the spectral
features and the stability of radial velocity measurements over a much
longer baseline.  In \S 2, we present our observations and the data
reduction.  In \S 3 we present the results from the spectral analysis
and the radial velocity study, including the P Cygni profiles and
transient high-velocity emission lines, which we interpret as emission
from bipolar jets. Similar high-velocity lines have been seen in
recent infrared spectra reported by Quaintrell \& Fender (1998).  In
\S 4 we discuss our attempts to model the complete binary system using
the constraints of these and previously published data on RXJ0019.

\section{Observations \& Data Reduction}

We monitored RXJ0019 at the Michigan-Dartmouth-M.I.T. (MDM)
observatory on Kitt Peak between 1994 December and 1997 February.
Most observations were made with the \telone telescope, the Mark III
spectrograph, a 600 {\it l}/mm grism, and a CCD detector.  In most
cases, our instrument configuration yielded a pixel size of 1.4 \AA\ 
and a resolution of ${\sim}3$\AA\ (FWHM). On occasion we used a
different CCD with wider pixels and a resolution (FWHM) of
${\sim}5$\AA\ (see Table ~\ref{tab:0019obs}). The 1995 November
observing run was made on the \teltwo telescope with similar
instrumentation and ${\sim}3$\AA\ resolution (FWHM).  In all, we
acquired 757 spectra of RXJ0019 with a total exposure of 180 ks.
Table ~\ref{tab:0019obs} summarizes the MDM observations.

We also monitored the source at the Fred Lawrence Whipple Observatory
(FLWO) located on Mt. Hopkins, AZ between 1994 July and 1995 February.
We used the \telct telescope and the FAST spectrograph (\cite{fab94})
and a pair of 1200 {\it l}/mm gratings.  The twenty-five FLWO
observations, which are listed in Table ~\ref{tab:0019obscfa}, were of
longer duration and of higher spectral resolution (${\sim}2$\AA\ FWHM)
than the MDM observations. To maintain a moderate range of wavelength
coverage, these data were obtained in alternating portions, one red
(5750-6700 \AA) and one blue (4550-5500 \AA).

Given that the orbital period of RXJ0019 is 15.85 hr, it is not
possible to observe one complete cycle in a night.  Furthermore, as
the period of RXJ0019 is in a near 2:3 resonance with the day/night
cycle, it is difficult to obtain uniform phase coverage. However,
these shortcomings are partially circumvented by our 2.5 year
observing campaign. During a typical MDM observing run, we pointed at
RXJ0019 several times per night (roughly every two hours), and we
obtained five 200-s exposures per pointing. Occasionally we continued
to take 200-s exposures of the target for several hours to search for
effects such as non-orbital fluctuations in the velocities of the
emission lines.

We used standard IRAF routines to reduce the raw CCD images into
one-dimensional extracted spectra and also to calibrate the spectra in
wavelength and flux.  In performing the wavelength calibrations, we
used the lamp spectra taken immediately before and after each pointing
sequence to interpolate the dispersion solution to the mid-exposure
time of a given observation.

\section{Analysis and Results}

\subsection{Radial Velocity and Period Determination}
\label{sec:0019rv}

Radial velocity measurements of RXJ0019 were extracted with IRAF tools
that compute cross correlations between two continuum-subtracted
spectra. In practice, the first of these spectra is a ``template'',
which is selected to have very nearly the same spectral features as
the object spectra. The velocity of an object spectrum, relative to a
restframe which is defined by the template, is determined by
cross-correlating the two spectra. All of the MDM observations of
RXJ0019 (1.4\AA/pixel) were cross-correlated with a template spectrum
constructed from the average of the 1996 January observations. Because
there is P Cygni absorption associated with the Balmer lines (see
below), the cross correlations were limited to wavelength intervals
selected for He\,II lines located at $\lambda \lambda$4200, 4542,
4686, 5412, and the series in the range of 6004$-$6407 \AA. A similar
analysis was conducted with the FLWO data, using the mean FLWO
spectrum as the cross correlation template.  Finally, prior to radial
velocity determinations, each of the two template spectra was
artificially redshifted so that the He\,II line centers were maximally
consistent with their corresponding laboratory wavelengths. The He\,II
line profiles in RXJ0019 are relatively narrow and symmetric, and we
do not expect significant systematic errors to arise from this
procedure.

For the MDM spectra, the radial velocities were computed and then
screened for discrepant values by flagging results that are
statistically inconsistent ($\geq 3 \sigma$) from the other 4
measurements obtained during the same telescope pointing. Only 19
spectra were eliminated for this reason.  Additionally, one set of
observations (December 1995b), produced unusually large variances
within groups of velocity measurements, and we eliminated these 60
spectra from further analyses. Thus, we derived a total of 691
velocity measurements from the two observatories. We used the
mid-exposure time of each observation to convert the velocities to the
heliocentric rest frame before proceeding with the timing investigations.

The binary period search was performed with a Stellingwerf dispersion
minimization routine (\cite{periods:stel}).  The variance statistic,
$\Theta$, is essentially the ratio of the average variance computed in
small phase bins to the variance of the data as a whole. When the data
are randomly distributed, $\Theta$ is approximately unity; however, if
the data become ordered, as when folded at the orbital period, 
$\Theta$ is reduced significantly.

The 691 velocity measurements were included in the search for the
period that minimizes $\Theta$.  The search covered trial periods in
the range of 8 - 20 hr. The results are shown in
Figure~\ref{fig:0019stel}, and the ${\sim}15.85$ hr period is easily
identifiable. The other two highly significant features are related to
the orbital frequency (${\sim}1.514$ cycles per day): the dip at 9.5
hr corresponds to the 1-day alias period (or 2.514 cycles per day),
while the dip at 19.1 hr corresponds to a frequency that is one-half
of the one-day alias.

We computed a least-squares parabolic fit for the values of $\Theta$
to localize the minimum near 15.85 hr. The results are shown in
Figure~\ref{fig:0019stelfit}, and the derived spectroscopic period and
uncertainty are included in Table ~\ref{tab:ephtable}.  We folded the
radial velocities about the spectroscopic period, as shown in
Figure~\ref{fig:0019rv}, and then we fit the results to a sine
function plus a constant to determine the best values for the velocity
semi-amplitude, the systemic velocity for the binary system, and the
time of maximum velocity.  The best values for the binary parameters
are given in Table ~\ref{tab:ephtable}.

Our spectroscopic period is consistent with the comparably accurate
photometric period determined by Will \& Barwig (1996; see Table
~\ref{tab:ephtable}). However, there is a discrepancy regarding the
relative phases of the velocity and photometric modulations. According
to Will \& Barwig (1996), a primary minimum should occur at HJD
$2450014.273 \pm 0.007$, which corresponds to a spectroscopic phase of
$0.171 \pm 0.013$. However, if one assumes that the binary orbit is
circular and that the accretion disk is circularly symmetric, then the
photometric minimum should occur at the time of superior conjunction
of the accreting star. This conjunction corresponds to a spectroscopic
phase of 0.25.  The phase offset between spectroscopic and photometric
results is therefore $0.079 \pm 0.013$, a discrepancy with a
significance of 6$\sigma$. Using data from 1992 August, Beuermann et
al. (1995) also reported that photometric minimum occurs 0.07 in phase
earlier than one would expect from the spectroscopic ephemeris, 
so this phase offset appears to be a stable characteristic of the
binary system.

\subsection{Deep Optical Spectrum}

Our extended monitoring provides not only radial velocity
measurements, but also about 160 ks of selected exposure at a
resolution of ${\sim}3$\AA.  While peculiarities in the spectrum of
RXJ0019, \eg, the extremely large He\,II($\lambda$4686)/H$\beta$
ratio, have been known since its discovery, the high signal-to-noise
of this new data set allows us to discern previously unreported
spectral features.

The different instrumental configurations have different spectral
bandpasses, but a substantial subset of the MDM observations has a
common region of overlap in the range of 4280-7000\AA\ (see Table
~\ref{tab:0019obs}). We constructed a ``deep rest frame'' spectrum
using the 398 observations obtained during 1995 June, 1995 December,
1996 January, and 1996 December. Each of the flux-calibrated spectra
was corrected for the Doppler motion predicted by the spectroscopic
ephemerides (Table ~\ref{tab:ephtable}), and the average of these
results is the rest frame spectrum shown in the top panel of 
Figure~\ref{fig:0019spec}. The P Cygni absorption features at H$\beta$
and H$\gamma$ are clearly evident, as is the strength of the He\,II
lines (e.g., He\,II $\lambda$4686) relative to lines of He\,I or H.

To enhance the visibility of subtle features in the rest frame
spectrum, we fitted the continuum to a smooth function and then
subtracted the continuum from the rest frame flux densities. The
results are shown in the bottom panel of Figure~\ref{fig:0019spec}
for the wavelength interval of 5600-6450\AA. This spectrum shows the
$n{\rightarrow}5$ transition series in He\,II, while the stronger
He\,II lines in Figure~\ref{fig:0019spec} arise from $n{\rightarrow}3$
and $n{\rightarrow}4$ transition series. To our knowledge, this is the
most detailed set of He\,II features observed in any celestial
object. In Table~\ref{tab:0019lines} we list all of the emission lines
detected in the rest frame spectrum, along with the line
identifications (when known) and the measured equivalent widths. The
high ionization levels represented by these spectral features are very
likely caused by photoionization related to the luminous, supersoft
X-ray emission in this system.
 
\subsection{Transient Jets}

In addition to the strong emission lines and P Cygni Balmer profiles,
some of the spectra of RXJ0019 exhibit red-shifted and blue-shifted
pairs of high-velocity lines. One such example, shown in
Figure~\ref{fig:cfa-outflow}, is the red and blue sequence of 900 s
exposures obtained at FLWO on 1994 December 12. The high velocity
lines are clearly separated from the central line, and we interpret
these results as the discovery of transient episodes of bipolar
outflow in RXJ0019.  High-velocity line pairs are evident at
H$\alpha$, H$\beta$, and He\,II $\lambda$4686 with approximate
equivalent widths of 3.0, 0.5, 1.0\AA, respectively.  These lines are
transient on time scales of months.  We also observed them during the
MDM observing runs of 1994 December, 1996 May, and 1996 December,
while they were not detected during the remaining observing runs.
RXJ0019 is now the second supersoft X-ray source to show high-velocity
emission lines. The LMC source RX J0513.9-6951 shows weak pairs of
emission lines near He\,II $\lambda$4686 and H$\beta$ corresponding to
an outflow velocity of approximately of 3900 km \ps\
(\cite{0513-6951:spect}; \cite{southwell}).

To track the radial velocities of all the spectral components related
to H$\alpha$, we first subtracted the continuum from the spectra
within ${\pm}4000$ km \ps\ of the line position. We then applied a
least-squares fit to the line profiles, using Gaussian functions to
represent, whenever present, the central line, P Cygni trough, red
outflow, and blue outflow components. In Figure~\ref{fig:satphase} we
plot the centroids of the Gaussian fits to the central and
high-velocity lines as functions of the orbital phase for H$\alpha$.
The radial velocities of these three components move in unison. The
results of chi-squared fits to these curves are given in
Table~\ref{tab:ha_velfits}.  All three $H\alpha$ lines are in phase
with the He\,II velocity curve to within statistical uncertainties.
Furthermore, we find that the $K$ amplitudes of the central, red, and
blue lines of H$\alpha$ agree with the $K$-velocity of the He\,II
$\lambda$4686 line to within 2.0$\sigma$, 0.1$\sigma$, and
1.6$\sigma$, respectively (see Table~\ref{tab:ha_velfits}).

The presence of P Cygni absorption (and any phase dependence in its
strength) will introduce systematic errors into the systemic velocity
(and $K$-velocity) values derived for the central component and the
blue outflow line.  Both of these effects are seen in
Table~\ref{tab:ha_velfits}. On the other hand, the red outflow line,
unaffected by P Cygni absorption, exhibits a velocity curve in
excellent agreement with the He\,II results. After correction for the
systemic velocity of --57.5 km \ps\ determined from the He\,II lines,
the red outflow line has a mean velocity of $815.5 \pm 10.0$ km \ps,
while the blue outflow line has a mean velocity of $-851.9 \pm 14.9$
km \ps, which is predictably shifted downward by P Cygni absorption
associated with the central component. We conclude that the jet 
lines are consistent with a symmetric model in which the outflow
velocity is 815 km \ps.

We note that the central and outflow lines at H$\alpha$ are all
significantly broader than the spectral resolution of the FLWO
instrument ($\sim$90 km \ps\ at H$\alpha$). We measure an intrinsic
width (FWHM) of about 400 km \ps\ for the jet lines.  This information
is used below to constrain geometric properties of the bipolar outflow
in RXJ0019.

Finally, in Figure~\ref{fig:pcyg_per} we plot the equivalent width of
P Cygni absorption versus orbital phase. There is a clear variation in
the strength of the absorption feature as the binary system
rotates. In particular, there is an absence of P Cygni absorption from
spectroscopic phase 0.8 to about 1.0, which corresponds to the phase
quadrant that begins just after the white dwarf passes closest to the
observer and ends when the two stars are aligned perpendicular to the
line of sight. The orbital phase dependence accounts for much of the
hour-to-hour variation in the P Cygni profile noted above. However,
the scatter in Figure~\ref{fig:pcyg_per} indicates that there is still
a significant random component to the absorption column. We return to
a discussion and interpretation of the P Cygni profiles in \S 4.

\section{System Modeling}

In this section, we focus on four specific results of the optical spectral
analysis to deduce additional binary system parameters and to
construct a simple kinematic model for the outflow of material
from the vicinity of the accretion disk.  The spectral results that we
utilize are summarized below:
 
\begin{enumerate}
 
\item The Doppler velocity curves for He\,II ($\lambda$4686),
H$\alpha$, and the red and blue jet lines of H$\alpha$ are in phase
and have a common $K$ amplitude of 71 km \ps. One expects the high
excitation He\,II lines, the central H$\alpha$ line, and the jet lines
to all originate in different physical regions. The most reasonable
interpretation for their common $K$-velocity is that it reflects the
orbital motion of the compact object.

The literature on cataclysmic variables contains many warnings that
the velocity curves derived from H emission lines do not yield
reliable dynamical results (e.g., \cite{shaft}).  In such cases the
line profiles indicate a complexity of emission-line sources, and the
derived $K$ velocities depend on the manner in which the cross
correlations are computed. Nevertheless, in RXJ0019 there are strong
arguments in favor of the interpretation that the measured
$K$-velocity does represent the orbital motion of the white dwarf:
First, our exclusive use of He\,II lines minimizes contamination by
low-excitation emission components far from the white dwarf. Secondly,
we see no evidence of systematic changes in the K velocity over a 2.5
year time scale; moreover, our results are also consistent with the
earlier work of Beuermann et al. (1995). Finally, as noted above, we
would not expect the central and outflow velocities to exhibit the
same velocity modulations unless there were a common origin for their
motion.  Our interpretation should, of course, be reassessed as future
optical measurements are made available.
 
\item The red and blue satellite lines, when present, have mean
projected velocities of 815 km \ps\ in opposite directions.
 
\item The  intrinsic width of these outflow lines is $\sim$400 km \ps.
 
\item P Cygni profiles are associated only with the Balmer lines, and
are most pronounced on the H$\beta$ and higher transitions.  The
strength of the P Cygni profiles varies systematically with orbital
phase.
 
\end{enumerate}
 
\subsection{Monte Carlo Constraints on Binary System Parameters}
\label{sec:montecarlo}
 
We utilize a model-dependent Monte Carlo method for estimating a
number of the binary system parameters for RXJ0019, starting with only
the known orbital period ($P$) and the $K$ amplitude ($71.2 \pm 3.6$
km \ps) of the accreting compact object, which is presumed to be a
white dwarf. The mass function can then be expressed as:
\begin{equation}
\label{eqn:massfunc1}
f(m) = \frac{M_{donor} \sin^3i}{\left(1+\frac{M_{wd}}{M_{donor}}\right)^2}
= \frac{K^3 P}{2 \pi G} = 0.0247 \pm 0.0040 \mbox{\msun.}
\end{equation}
It is apparent from the small value of the mass function that in order
for the donor star to have a plausible mass, either the inclination
angle must be small, or the mass of the accretor must be large
compared to that of the donor.
 
If we utilize a few model-dependent ideas about how an accreting
binary such as RXJ0019 may evolve, we can combine the measured mass
function with a Monte Carlo sampling of evolving binaries to infer a
substantial number of system parameters. Such an approach was
developed previously to infer the system parameters for the
binary X-ray pulsar GROJ1744-28 (\cite {gro1744:rj97}). We briefly
outline that method here.

We start with a short period binary system consisting of a donor star
and a white dwarf with initial masses in the following ranges:
\begin{eqnarray}
\label{eqn:mloss}
0.8 \leq & M_{donor}/M_{\odot} & \leq 3.0\\
0.6 \leq  & M_{wd}/M_{\odot} & \leq 1.2\;.
\end{eqnarray}
We begin the binary evolution calculation only after the formation of
the white dwarf.  This occurs at the end point of a common envelope
phase in which the giant progenitor of the white dwarf has its
envelope stripped off by the companion star (the ``donor'' in the
supersoft X-ray source phase). The lower limit of $\sim$0.8 \msun\ for
the mass of the donor star is based on the requirement that the binary
system eventually reaches a period of 15.85 hr. Donor stars with lower
initial mass will come into Roche-lobe contact while on the main
sequence with binary periods much shorter than 15 hr and will evolve
toward shorter orbital periods (except in the case of extreme X-ray
heating effects; see discussion below). In initially wider orbits,
such low-mass donors would have insufficient time within the age of
the Galaxy to evolve and fill their Roche lobe. The lower limit for the
mass of the white dwarf (0.6 \msun) is required to attain the high
surface temperatures observed in supersoft X-ray sources
(\cite{burning:iben82}; HBNR; RDS). Finally, the upper limit of 1.2
\msun\ for the mass of the white dwarf is taken from the population
synthesis study of RDS.

Our approach is to choose, via a Monte Carlo method, the post-common
envelope properties of the binary system, evolve the binary to the
point where its orbital period reaches 15.85 hr, and then find the
orbital inclination angle that will yield the observed $K$ velocity of
the white dwarf.  If the orbital period never reaches 15.85 hr, then
the system is discarded. If the system does reach a state in which it
could be a plausible match for RXJ0019 (i.e., with a 15.85 hr period
and an orbital velocity for the white dwarf at least as large as the
observed $K$-velocity), then the system parameters at that point are
saved.  The cumulative statistics for all the binaries that produce
RXJ0019-like systems form probability distributions for the orbital
inclination angle, donor mass, core mass of the donor, and the
luminosity of the donor. We may then impose further constraints on the
RXJ0019-like systems, such as the mass transfer rate. This yields
probability distributions for the system parameters that are
consistent with the mass transfer rate being sufficiently high to
support steady (or quasi-steady) nuclear burning on the white dwarf,
$\sim$ \scinot{3}{-8} $< \dot{M} < 10^{-6}$ \mspyr\
(\cite{burning:iben82}; HBNR; RDS).

In principle, the best way to choose the initial (\ie, immediately
after the common envelope phase) binary system parameters is from the
output of a population synthesis code (see, e.g., RDS). However, for
purposes of this exercise it is sufficient to choose the masses of the
white dwarf and the donor star with uniform probability over the
ranges that are permitted. We must also specify the initial core mass,
$M_c$, of the donor star.  For simplicity, we pick the core mass from
a uniform distribution over the range $0 < M_c < 0.15$ \msun, above
which point the orbital period of the system would always be longer
than 15.85 hr. The evolution is started with the stars separated by a
distance which corresponds to the secondary star just filling its
Roche lobe. Further details about the binary evolution code are given
by Di\,Stefano \etal\ (1997).

This procedure of choosing initial binary parameters and then evolving
the system to see if it would become a near match to RXJ0019 was
carried out \scinot{2}{6} times. We recorded the binary
parameters for 437,000 successful systems.  In each case the results
were weighted by a factor of $\sin^2i/\cos i$, which takes into
account the probability of observing randomly oriented systems with
the correct inclination angle to produce the observed mass function
(see \cite{gro1744:rj97}). The results are shown in
Figure~\ref{fig:incline_hist} where probability histograms are given
for the inclination angle and mass of the donor star.  The current
value of the binary period of RXJ0019 can be approached from either
longer periods (with $\dot{P} < 0$), or after the binary has passed
through its minimum orbital period (with $\dot{P} > 0$), and we keep
track of these cases separately. Each histogram in
Figure~\ref{fig:incline_hist} also displays shaded regions that
indicate those binaries with moderately high rates of mass transfer
(\scinot{3}{-8} \mspyr\ $< \dot{M} <$ \tento{-7} \mspyr;
single hatching) and those with very high rates of mass transfer
($\dot{M} > $ \tento{-7}\mspyr; double hatching). The unhatched
histograms include all mass transfer rates.

From this figure we learn that the values of binary period and the
emission-line $K$- velocity, combined with acceptable values for the
initial stellar masses, constrain the inclination angle to the range
of $16^\circ - 40^\circ$, regardless of the mass transfer rate.
Furthermore, if we assume that there is at least a modest rate of mass
transfer ($\dot{M} >$ \scinot{3}{-8} \mspyr), then the inclination
angle is {\em un}likely to be greater than 30$^\circ$. Both of these
inclination limits can be avoided only if (i) the $K$ velocity that we
have used is not the correct one to be associated with the orbital
motion of the accreting star, or (ii) one allows for more extreme
constituent masses, \eg, a 1.5 \msun\ donor star orbiting an 8 \msun\
black hole, or a 0.5 \msun\ donor star orbiting a 1.3 \msun\ white
dwarf --- both of which would yield an inclination angle of
$\sim60^\circ$. With regard to point (ii), it is possible that some
evolutionary scenario other than the one utilized in our Monte Carlo
study can explain persistent supersoft X-ray sources. We note that van
Teeseling \& King (1998) have recently proposed a model wherein the
mass transfer and binary evolution in some supersoft X-ray sources are
driven by wind loss from an X-ray irradiated, low-mass donor star.

One difficulty with a small inclination angle arises in the effort to
explain the photometric light curve (\cite{0019:photometry}).  This
curve shows a highly repeatable modulation at the binary period with a
full depth of 0.5 mag.  It seems difficult to explain such behavior
with inclination angles as small as $20^\circ - 30^\circ$ (see, \eg,
\cite{sxsdisk:smm97}). Alternatively, the results could signify
substantial vertical structure in the accretion disks of supersoft
X-ray sources, perhaps as a consequence of the high rate of mass
transfer. Some hydrodynamic disk simulations do show impressive
structures with significant vertical extent (\cite{diskmod}).

\subsection{Kinematic Model for the Jet Lines}
 
        We now examine more closely at least one possible
interpretation of the jet lines.  In this study, we interpret these
lines as representing material flowing away from the accretion disk.
In order to constrain the geometry of the jet, we adopt a very simple
model for the outflow, which is based on spectral results (1) and (2)
in \S 4.0. We define a jet cone into which the emitting atoms are
released, all with speed $v_j$. The cone is defined by a tilt angle,
$\theta_j$ (with respect to the angular momentum axis of the binary),
an opening angle of half-width, $\alpha$, and an azimuthal angle,
$\phi_j$.  Note that $\theta_j$ and $\phi_j$ refer to the symmetry
axis of the jet. Material is assumed to come off uniformly (per unit
solid angle) within the cone during the times when the outflow lines
are visible. There are jet cones located on either side of the
accretion disk. In principle, the four jet parameters need not have
identical values above and below the disk, and ideally we should fit
for these parameters independently. However, since the blue-shifted
line is unfortunately contaminated by P Cygni absorption, we focus our
attention on the red-shifted portion of the outflow.

At least five effects contribute to the observed profiles of the jet
lines. These include the Doppler shifts projected along the line of
sight due to (1) the velocity of the outflowing matter along the
symmetry axis of the jet cone; (2) the orbital motion of the whole
accretion disk around the center of mass in the binary system; (3) the
different directions of atoms in the jet with respect to the axis of
the cone; (4) the Keplerian motion of matter in the accretion disk
around the white dwarf; and (5) the height above the disk of the jet
emission region.  The first two of these tend to dominate the position
of the center of the line, while the next two tend to broaden the line
and perhaps induce asymmetries. The fifth factor influences the
detectability of binary motion in the velocities of the jet lines. We
first examine how the centroids of the lines vary with time around the
orbit under the assumption that the peak of the line is governed by
(1) and (2) only. We later discuss calculations of the full line
profile which justify this assumption, and we indicate what limits can
be placed on effects (3), (4), and (5).
 
With no loss of generality, we consider the case in which the binary
orbit lies in the $xy$ plane with the observer located in the $xz$
plane at an inclination angle, $i$, with respect to the $z$ axis.  Let
$\vec{k}$, $\vec{V_o}$, and $\vec{V_j}$ be, respectively, the unit
vector pointing from the observer to the binary, the orbital velocity
of the white dwarf, and the velocity of an atom moving along the jet
axis with speed $v_j$ in a frame orbiting with the white dwarf. In terms
of their Cartesian coordinates, these vectors can be written as:
\begin{eqnarray}
\label{eqn:los}
&\vec{k} =    \left[\:-\sin(i),\; 0,\;  -\cos(i)\:\right],&\\
&\vec{V_o} = v_o \left[\:-\cos(\omega t),\; -\sin(\omega t),\; 0 \:\right],&\\
\label{eqn:velspray}
&\vec{V_j} = v_j \left[\:\sin(\theta_j) \cos(\phi_j),\; \sin(\theta_j) \sin(\phi_j),\; \cos(\theta_j) \:\right],&
\end{eqnarray}
where $\omega$ is the Keplerian angular frequency of the binary, and
$\theta_j$ and $\phi_j$ are the instantaneous spherical polar angles 
describing the orientatiuon of the jet axis. With these definitions,
and additional consideration of the systemic radial velocity
($\gamma$) of the binary system, one can predict the observed Doppler
velocity for the line centroids due to effects (1) and (2) above. If
we define $V_C$ and $V_{out}$ as the Doppler velocities for the central
and jet outflow lines, respectively, we find:
\begin{eqnarray}
\label{eqn:jetvel_line}
&V_{C} =   \gamma + v_o \sin(i) \cos(\omega t) \equiv \gamma + K \cos(\omega t),&\\
\label{eqn:jetvel_red}
&V_{out} =   \gamma + K \cos(\omega t) - v_j \sin(i) \sin(\theta_j)
\cos(\phi_j) - v_j \cos(i) \cos(\theta_j).&
\end{eqnarray} 

We can now utilize our radial velocity measurements for RXJ0019 to
extract the parameters associated with the outflow. In order to
proceed, we consider two possibilities: ($case$ $a$) the jet axis is
tilted by an angle $\theta_j$ with respect to the normal to the
accretion disk (which is assumed to lie in the orbital plane), and its
projection into the orbital plane is fixed with respect to the line
joining the two stars; and ($case$ $b$) the jet axis is normal to the
accretion disk, which in turn is tilted with respect to the orbital
plane by an angle $\theta_j$.  In both cases, the angle $\theta_j$ is
taken to be a constant; however, in case (a) the azimuth angle,
$\phi_j$, varies with the orbital period (i.e., $\phi_j = \omega t$ +
constant), while in case (b) $\phi_j$ will vary with the disk
precession time which, in general, should be much longer than the
orbital period.
 
In case (a) we can extract a rather tight constraint on $\theta_j$ as
follows. As shown above, the velocities of the central H$\alpha$ line
are consistent with the best fit for the He\,II lines, i.e. $V_C =
V_{HeII}$.  Therefore, we may effectively remove the first two terms
in equation (\ref{eqn:jetvel_red}) by subtracting $V_{HeII}$ from the
observed Doppler velocities for the red jet line of H$\alpha$, i.e.,
$V_{out}-V_{HeII}$. We then compute the average of the subtracted
results, which we associate with the last term in equation
(\ref{eqn:jetvel_red}), i.e.:
\begin{equation}
\label{eqn:velavg} 
\langle V_{out} - V_{HeII} \rangle = v_j \cos(i) \cos(\theta_j) = 815\;\; \mbox{km \ps.}
\end{equation} 

Finally, we isolate the third term in equation (\ref{eqn:jetvel_red})
by first subtracting both $V_{HeII}$ and the result of equation
(\ref{eqn:velavg}) from equation (\ref{eqn:jetvel_red}).  We then 
fit a sine curve to the residuals of this operation. The result is:
\begin{equation}
\label{eqn:sindet} 
 v_j \sin(i) \sin(\theta_j) = 1.8 {\pm} 14\;\; \mbox{km \ps\ [case (a)],}
\end{equation} 
for the red outflow line, where the fit was made to the data already
folded at the orbital period.

From the above analysis, we can set a combined limit on the
inclination angle and the jet tilt angle by dividing
equation (\ref{eqn:sindet}) by (\ref{eqn:velavg}) to obtain, 
at 90\% confidence:
\begin{equation}
\label{eqn:tandet} 
\tan(i) \tan(\theta_j)  <  0.027\;\;  \mbox{[case (a)],}
\end{equation}  
for the red jet.  If we now utilize the lower limit of $\sim 16 ^\circ$
for the orbital inclination angle, we can set a limit (again at 
90\% confidence) on the tilt angle:
\begin{equation}
\label{eqn:tilt} 
\theta_j   <  5^\circ.4 \;\;  \mbox{[case (a)].}
\end{equation}

Returning now to case (b), where the azimuthal angle $\phi_j$ is
assumed to vary only on a disk precession timescale, we analyze the
radial velocities of the red outflow line in an analogous way to that
just described.  In this case, we assume only that the putative disk
precession period is less than $\sim$300 days.  (For substantially
longer precession periods, the only constraint that we can set on
$\theta_j$, $\phi_j$, $v_j$, and $i$ is that the sum of the last two
terms in equation (\ref{eqn:jetvel_red}) equals 815 km \ps\ during the
epoch of our observations.)  For precession periods less than
$\sim$300 days (i.e., where $\phi_j$ has moved through an angle much
greater than 2$\pi$ over the interval of our observations), we can
still isolate the middle term in equation (\ref{eqn:jetvel_red}) to
set constraints on $\theta_j$ and $i$. Again, we subtracted $V_{HeII}$
from the time series of red outflow velocities. We then carried out a
Stellingwerf period search on the residuals of this operation.  This
search covered possible disk precession periods ranging from 0.5 to
300 d.  From this analysis we derive a limit on the central term in
equation (\ref{eqn:jetvel_red}) of:
\begin{equation}
\label{eqn:sindetb} 
 v_j \sin(i) \sin(\theta_j) < 30\;\; \mbox{km \ps\ [case (b)],}\;\;
\end{equation} 
at the 90\% confidence level.  In this case, the limit analogous
to that given by equation (\ref{eqn:tandet}) is
\begin{equation}
\label{eqn:tandetb} 
\tan(i) \tan(\theta_j)  <  0.037\;\;  \mbox{[case (b)].}
\end{equation}  

Again, if we adopt a lower limit on the inclination angle of $16^\circ $, we
find a corresponding limit on the tilt angle of:
\begin{equation}
\label{eqn:tiltb} 
\theta_j  <  7.3^\circ \;\; \mbox{[case (b)].}
\end{equation}

Finally, as noted previously, the blue outflow line at H$\alpha$ is
blended with P Cygni absorption, and the
deconvolution of these features is beset with systematic problems
related to our limited spectral resolution and the uncertainties in
the specific line profiles for each component.  Nevertheless if we do
investigate the radial velocity measurements of the blue outflow line
for evidence of a tilt angle with respect to the binary plane, as we
have done for the red line, then the corresponding 90\% confidence
limits to equations (\ref{eqn:tilt}) and (\ref{eqn:tiltb}) are
$10^\circ $ and $15^\circ $, respectively, consistent with the results
for the red line.

From the above analyses, we conclude simply that the jet axes are
essentially perpendicular to the orbital plane of the binary.
 
\subsection{Emission-line Profiles}

In the context of our simple kinematic model, we now consider two
effects which would broaden the emission lines associated with the
outflow.  Briefly, these are (1) the finite opening (half) angle,
$\alpha$, of the jet, and (2) Keplerian motion of matter in the
accretion disk that may feed the jet.  We consider each of these in
turn, and we assume in the calculations below that the outflow
emission region is optically thin.
 
The opening angle of the jet affects both the width and the shape of
the outflow lines. In order to constrain this opening angle, we adopt
a simple model wherein the jet cone has a uniform flux of emitting
atoms per unit solid angle over the entire cone.  We then utilize
equations (\ref{eqn:los}) through (\ref{eqn:velspray}) to integrate
the contributions over atoms moving in different directions within the
outflow. However, equation (\ref{eqn:velspray}) must be generalized to
refer to the velocity of {\em any} atom within the outflow, and we
apply these considerations to a cone of atoms moving with a
component along the red jet axis. Since we concluded in the preceding
section that the jet cone axis is essentially normal to the orbital
plane, \ie, $\theta_j = 0$, equation (\ref{eqn:jetvel_red}) becomes:
\begin{equation}
\label{eqn:rb_atom}
V_{a} = \gamma + K \sin(\omega t) - v_j \sin(i) \sin(\theta) \cos(\phi) 
        + v_j \cos(i) \cos(\theta),
\end{equation}
where $V_a$ is the Doppler velocity of a single atom in the jet whose
direction is given by $\theta$ with respect to the red jet axis.  In
order to compute the line profile due to the range of angles within
the jet cone, one simply integrates the contributions from all atoms
with $\theta < \alpha$ (the half width of the opening angle), and $0 <
\phi < 2\pi$.  After Doppler correcting the spectra to the rest frame
of the accretion disk, one needs to sum only the contributions from
the last two terms in equation (\ref{eqn:rb_atom}) to find the line
profile due to the finite opening angle of the jet.
 
There is potentially a wealth of information in the profiles of the
jet lines, which we earlier determined to have an intrinsic FWHM of
$\sim400$ km \ps. In principle, the central wavelength, width,
asymmetry, and movement of the jet lines allow one to determine the
outflow velocity as well as the origin and opening angle of the
jet. However, given our modest instrumental resolution, we may only
derive coarse constraints on these parameters.

First, we calculated a grid of theoretical model profiles for various
values of the opening half-angle, $\alpha$, and the orbital
inclination, $i$.  For each grid point the jet line profile was
computed and then convolved with the instrumental resolution ($\sigma
=$ FWHM $/ 2.355 = 40$ km \ps\ for the FLWO data).  Next, the model
profiles were renormalized in flux and shifted in velocity to best
match the observed profile of the red jet line, using a chi-square fit
over the range of $6568-6620$ \AA. The red jet line was isolated by
fitting the central H$\alpha$ line to a Gaussian and then subtracting
the best fit from the spectrum. The line profile is distinctly
asymmetric, falling off more steeply toward longer wavelength, as do
the majority of the theoretical model profiles.  Finally, we
constructed contours of constant (minimum) chi square values in the
$\alpha$-$i$ plane. The results are shown in
Figure~\ref{fig:alpha_incl}. Because the model is so simplistic, and
after comparing the fits to the observations, we judge a reduced chi
square value below 5.0 to be a reasonable representation of the
observations; this contour is shown with a darkened line in Figure
~\ref{fig:alpha_incl}.  Our simple jet model prefers a binary
inclination angle in the range of $35^\circ - 60^\circ$ with a jet
opening half-angle in the range of $40^\circ \ga \alpha \ga 15^\circ$.

If these rough limits on the opening half-angle are correct, then the
jet outflows in RXJ0019 are not well collimated. Alternatively, the
jets could be well collimated, while the line widths are increased by
other effects, including a range of velocities within the line
formation region.  One particular additional cause of line broadening
is motion of the disk material orbiting the compact star.  The
Keplerian velocity at radial distance $r$ is
\begin{equation}
\label{eqn:veldisk} 
v_{disk}  =  \left(\frac{GM_{wd}}{r}\right)^{1/2}  =
1150 \left(\frac{M_{wd}}{M_\odot}\right)^{1/2} r_{10}^{-1/2}\;\;
\mbox{km \ps,}
\end{equation} 
where $r_{10}$ is the radial distance from the compact object (assumed
to be a white dwarf) in units of \tento{10} cm.  If the accretion disk
lies in the orbital plane and the system is viewed from an inclination
angle, $i$, then the maximum Doppler velocity due to matter orbiting
in the accretion disk in an annulus at radius $r$ is $\pm 1150\;
(M_{wd}/M_\odot)^{1/2} r_{10}^{-1/2} \sin(i)$ km \ps.  However, from
the summary of spectral facts cited at the beginning of \S4, we see
that the FWHM of the H$\alpha$ lines (central and outflow
lines) is only 400 km \ps.  Therefore, we conclude that the regions of
the accretion disk contributing to the H$\alpha$ emission are located
relatively far from the the white dwarf at radial distances
\begin{equation}
\label{eqn:radlimit}
r \ga 2.6{\times}10^{10} \left(\frac{M_{wd}}{M_\odot}\right)
 \left[\frac{\sin(i)}{\sin(25^\circ)}\right]^2\;\mbox{cm.}
\end{equation} 
For a white dwarf of 1 \msun\ and a radius of
$\sim$\scinot{5}{8} cm, the H$\alpha$ emission region must lie at
least $\sim$40 white dwarf radii from the compact star.  This suggests
that the 815 km \ps\ jet is not produced near the compact star, but
rather emanates from a region further out in the disk. The escape
velocity from a 1 \msun\ white dwarf is $\sim$7000 km \ps, so that the
observed projected outflow speed of 815 km \ps\ again suggests that
the ejection is taking place relatively far from the surface of the
white dwarf.
 
A final caveat in regard to possible broadening due to the Keplerian
motion of the disk material is in order.  If the outflow originates
very near the compact star, \eg, if the outflow were jet-like, and the
mechanism driving the outflow could remove the Keplerian motion of the
inner disk from the ejected matter, then perhaps the observed Doppler
velocities would not have any broadening due to the rotation of the
disk material. In this case, our limit for the inner radius of the
emission line region given by equation (\ref{eqn:radlimit}) might not
be valid.
 
Lastly, we comment on the fact that both the red and blue outflow
emission lines, when present, always have approximately the same
intensity. This implies that the emission region must be far enough
from the disk that the outflow on either side is equally visible. On
the other hand, the emission region cannot be so far from the disk
that the orbital Doppler effect that we clearly detect, would be
smeared out by emission originating over different orbital phases. 

The first requirement implies the obvious geometric constraint on $H$,
the height above the disk at which the emission takes place. We may
express the constraint on $H$ relative to the semimajor axis of the
orbit of the compact star, $a_x$:
\begin{equation} 
\label{eqn:geomcontr}
\frac{H}{a_x}  > \frac{R_D}{a_x \tan(i)}\;  
= \;R_D \frac{(1+q)}{a \; q \; \tan(i)}\;
=\; \frac{0.46 f}{ \tan(i)}\frac{(1+q)^{2/3}}{q}
\end{equation} 
where $R_D$ is the radius of the accretion disk, $q$ is the mass ratio
$M_{don}/M_{wd}$, and $f$ is the fraction of the Roche lobe of the compact
star that the accretion disk occupies. The approximation for $R_D/a$
in equation (\ref{eqn:geomcontr}) is accurate to within $\sim10$\% over
the range $q < 2$.  For typical system parameters and $i \simeq
25^\circ$ and $f = 0.7$, we expect $H/a_x \ga 1$.

The Doppler modulation of the jet lines at the binary period
constrains the upper limit for the emitting region above the accretion
disk. This limit can be expressed as:
\begin{equation}
\label{eqn:sheight}
H  <  v_j/\omega  = \frac{815 \; \mbox{km \ps}}{\omega \cos(i)}\;.
\end{equation} 
We can also cast this limit in dimensionless form by dividing by
$a_x$, the semimajor axis of the orbit of the compact star:
\begin{equation}
\label{eqn:ht_uplim} 
\frac{H}{a_x} < \frac{v_j}{\omega a_x} = \frac{v_j}{K/sin(i)} = 11.5 \tan(i)\;.
\end{equation}
For $i \simeq 25^\circ$, we find that $H/a_x < 5.3$.  

In summary, if the emitting region is no further above or below the accretion
disk than ${\sim}6\ a_x$, nor any closer than ${\sim}1\ a_x$, then both
the red and the blue jet lines should be approximately equally visible
while retaining the net Doppler motion at the binary period, as we
have observed.  
 
\section{P Cygni Profiles}
\label{sec:pcygni}

As shown in Figure~\ref{fig:pcyg_per} the strength of the P Cygni
absorption varies with orbital phase. The strength of the absorption
feature is a minimum at spectroscopic phases 0.8 to 1.0, when the
central line is moving from its rest wavelength position to its
maximum redshift.  A similar result was reported by \cite{0019:pcyg}. If
the shape of the P Cygni trough and its central wavelength are
independent of orbital phase, then the Doppler motion in the central
line could reduce the net absorption strength near phase 0.5. Since
that is nearly the opposite of what we observe, we conclude that the
phase dependence of the absorption strength is intrinsic to the
geometry of the absorbing gas.

The nature and origin of the gas responsible for the P Cygni
absorption remain a mystery.  Our first hypothesis was that the gas
responsible for the jet outflow might also be the gas that causes P
Cygni absorption. The phase dependence of P Cygni absorption implies
an opening half-angle $\sim 100^\circ$ for the absorbing gas which is
much larger than the $\sim 40^\circ$ upper limit for $\alpha$ in the jet.
Furthermore, the absorption cone is not only wide, but it is
azimuthally asymmetric, while the jet outflow has been shown to be
closely aligned with the binary spin axis. Finally, the jet and P
Cygni components appear to have differences in their persistence time
scales. Although the P Cygni absorption strength is variable, it is
apparent in all of the observing runs.  By contrast, the jet lines are
present only in about one-third of our observing epochs.  Therefore,
the very different behavior between the jet lines and the P Cygni
absorption features leads us to believe that they arise from different
physical regions in the binary system.
 
\section{Summary and Conclusions}
 
\begin{enumerate}
 
\item  We have made spectroscopic observations of RX J0019.8+2156 over
a 2.5 year baseline, collecting over 200 ks of data on this source.
 
\item Radial velocity studies of the He\,II $\lambda$4686\AA\ emission
line have yielded a Doppler curve with a velocity half-amplitude of
$71.2 \pm\ 3.6$ km \ps (see Table~\ref{tab:ephtable}). The binary mass
function is $0.0247 \pm 0.0040$ \msun.
 
\item The period as determined from our radial velocity study is in
agreement with the photometrically determined value obtained from the
literature.  However, the phases determined by the two methods
disagree by 28$^\circ$ ($6 \sigma$ significance), assuming that the
binary orbit is circular and the accretion disk is circularly
symmetric.
 
\item The summed rest-frame spectrum shows extraordinary He\,II series
emission.  Most notably, the $n{\rightarrow}5$ transition series is
visible for values of $n$ up to 22.
 
\item A model-dependent population synthesis study indicates that if
the binary parameters are to be consistent with the measured mass
function, then $16^\circ \la i \la 40^\circ$.  Furthermore, if the
system also has a mass ratio which is capable of driving high mass
transfer rates (in the context of the van den Heuvel \etal\ model),
then the orbital inclination angle is $16^\circ \la i \la 30^\circ$.
However, such a low inclination is difficult to reconcile with
published photometry, which shows a pronounced primary and secondary
dip.
 
\item We discovered the existence of low-velocity bipolar jets that
are revealed as red-shifted and blue-shifted emission lines that flank
H$\alpha$, H$\beta$, and He\,II $\lambda$4686.  These lines are
transient and were detected only during the 1994 Dec, 1996 May, and
1996 Dec. After correcting for the gamma velocity of the system,
these lines are nearly symmetrical around the central line with a
velocity of $\pm$ 815 km \ps.
 
\item The outflow lines also exhibit orbital Doppler motion with the
same amplitude and phase as that determined by the He\,II lines.  This
implies that the jets are oriented nearly perpendicular to the orbital
plane.

\item The profiles of the outflow lines potentially possess a great
deal of information. While the limited spectral resolution limits our
analysis, we are nevertheless able to conclude that the range of
inclination angles is likely to be $ 35^\circ \la i \la 60^\circ$, and
the corresponding range of opening half-angles $40^\circ \ga \alpha
\ga 15^\circ$, if the intrinsic width in the line profiles is
dominated by the opening geometry of the jet.

\item In order for both red and blue outflow lines to be observable at
all orbital phases and also exhibit the Doppler curve of the white
dwarf, the emission must originate from a distance $H$ from the disk
in the range $1 \la H / a_x \la 6$, where $a_x$ is the semimajor axis
of the orbit of the compact object.
 
\item We discovered that the strength of the P Cygni profiles is
variable with orbital phase. The $\sim100^\circ$ opening half-angle of
the absorbing gas and its asymmetry with respect to the line joining
the stars is distinctly different from the geometry of the gas
emanating from the jets.  For this reason, and because of the
transient nature of the outflow lines, it is unlikely that the
absorption and jet components have the same origin.
 
\end{enumerate}

Acknowledgements: We thank John Thorstensen, Jules Halpern, Mike
Eracleous, and Bob Barr for helping with the observations at MDM
Observatory. Jim Peters and Perry Berlind made observations for us at
FLWO, and Susan Tokarz helped with the reduction of FLWO spectra.
Travel support to RR and CB for these optical observations was
provided by NSF grant AST-9315074. This work was also supported, in
part, by NASA grant NAG5-3011.

\clearpage

% ---------------------------------------------------------------
% -------------------     Bibiliography               ----------------
% ---------------------------------------------------------------

\clearpage

\begin{center}
{\bf Figure Legends}
\end{center}

\figcaption[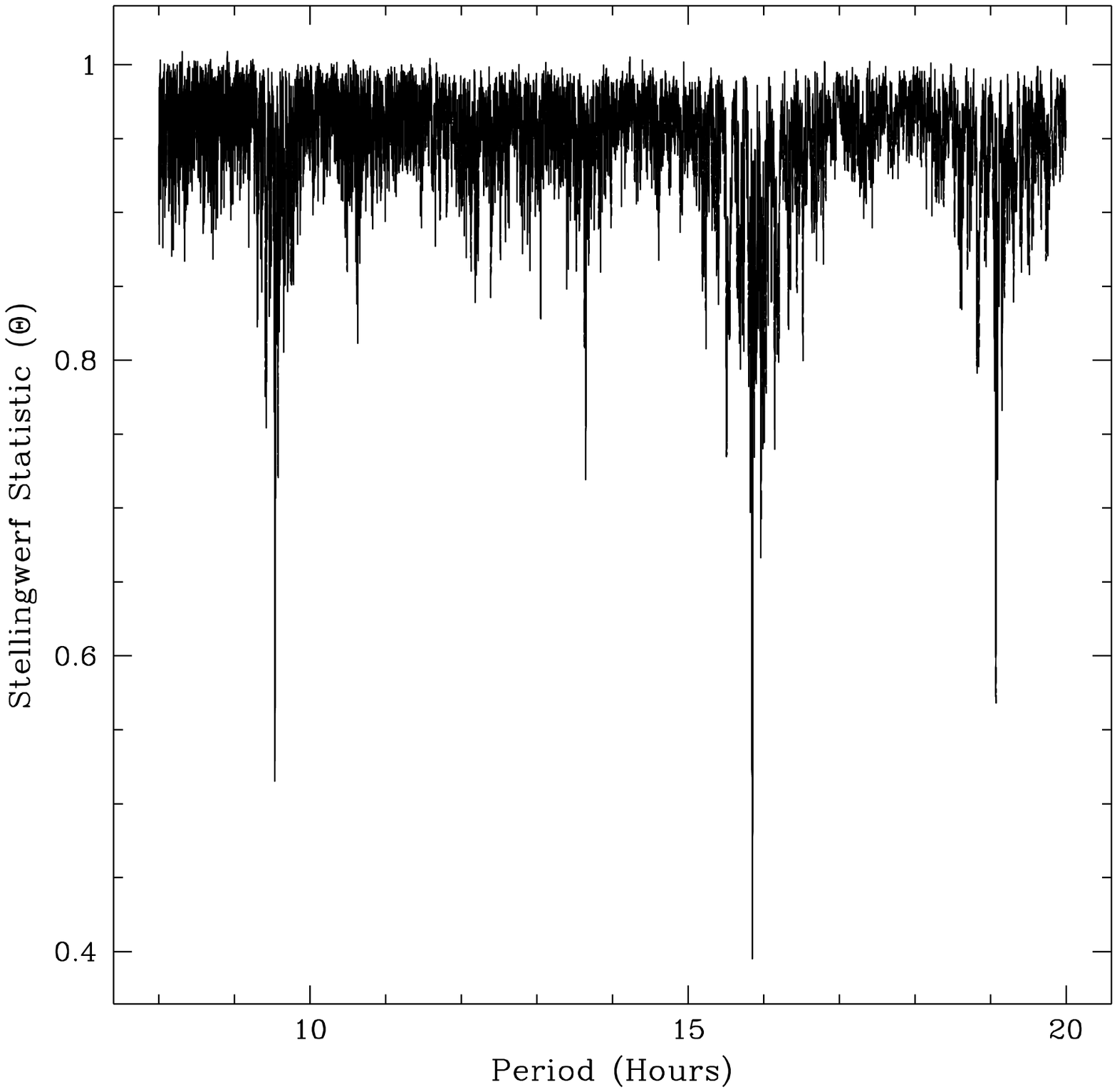]{Stellingwerf period search using radial velocity
measurements of RXJ0019. Minima in the variance statistic correspond
to periodicities in the data. \label{fig:0019stel}}

\figcaption[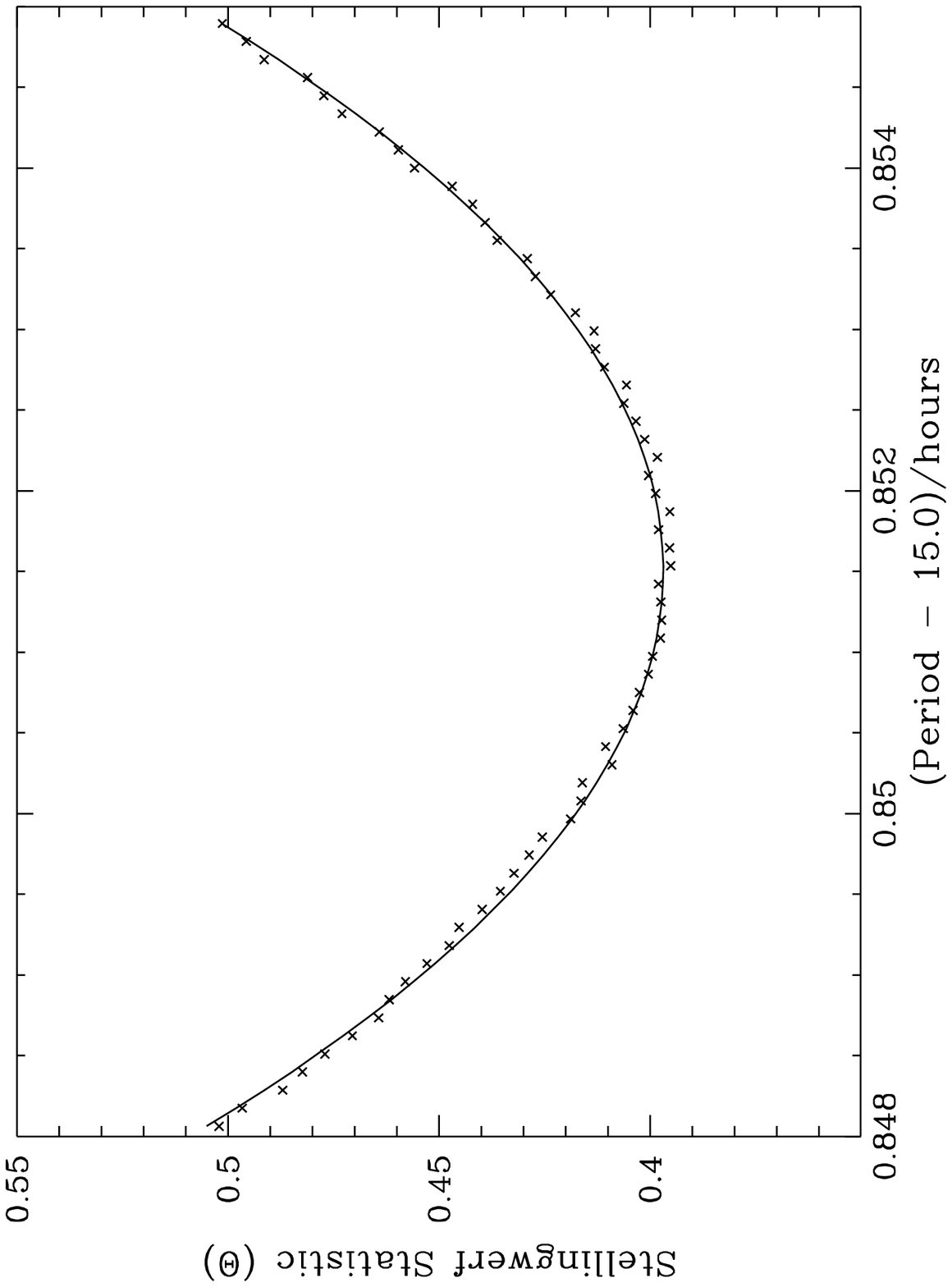]{Enlargement and fit to the minimum in the
variance statistic near a trial period of 15.85 hr. \label{fig:0019stelfit}}

\figcaption[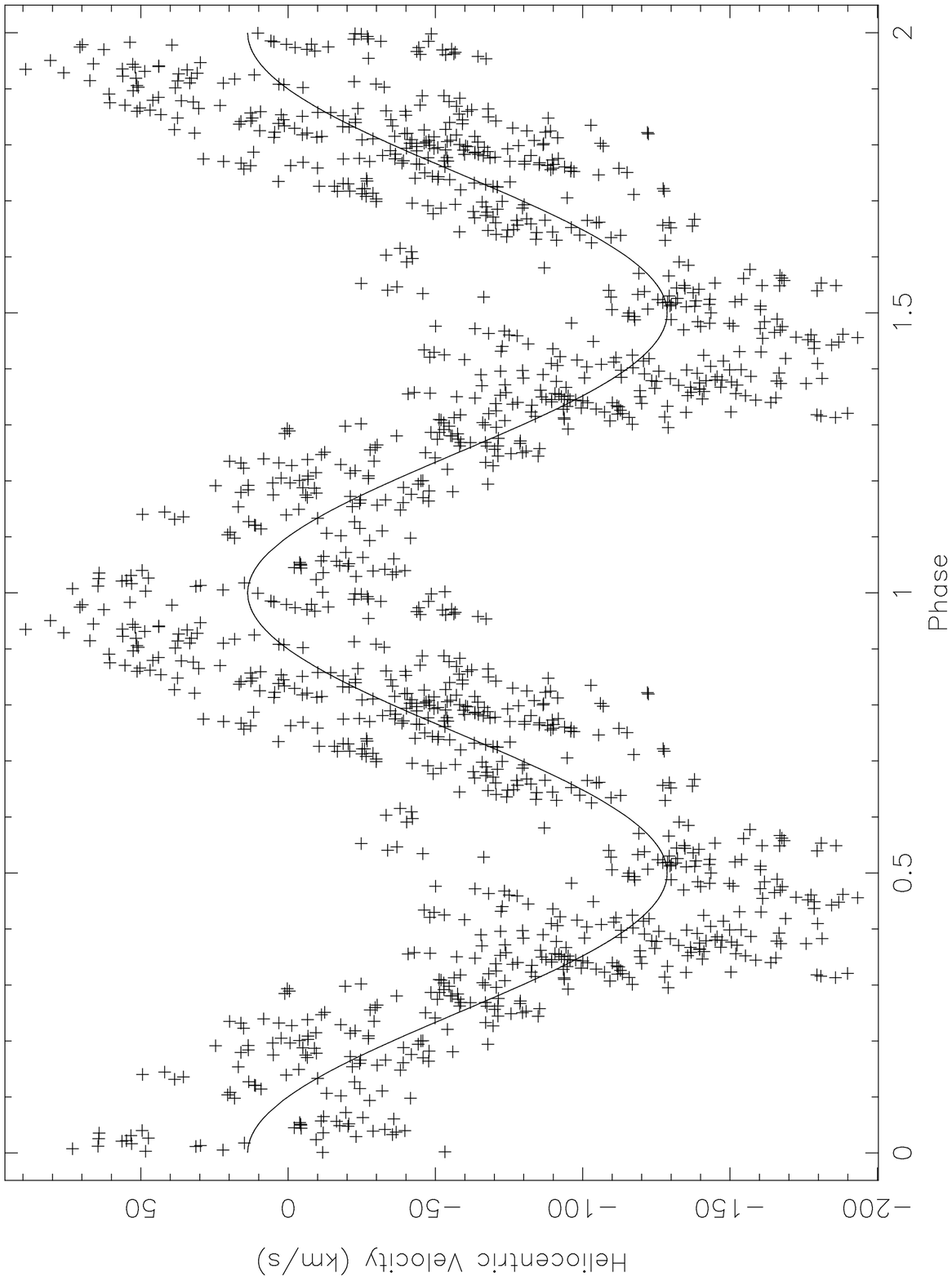]{Heliocentric velocities of the HeII line folded
about the 15.85151 hr period and a sine wave fit. The data span an
interval of $\sim$2.5 years. \label{fig:0019rv}}

\figcaption[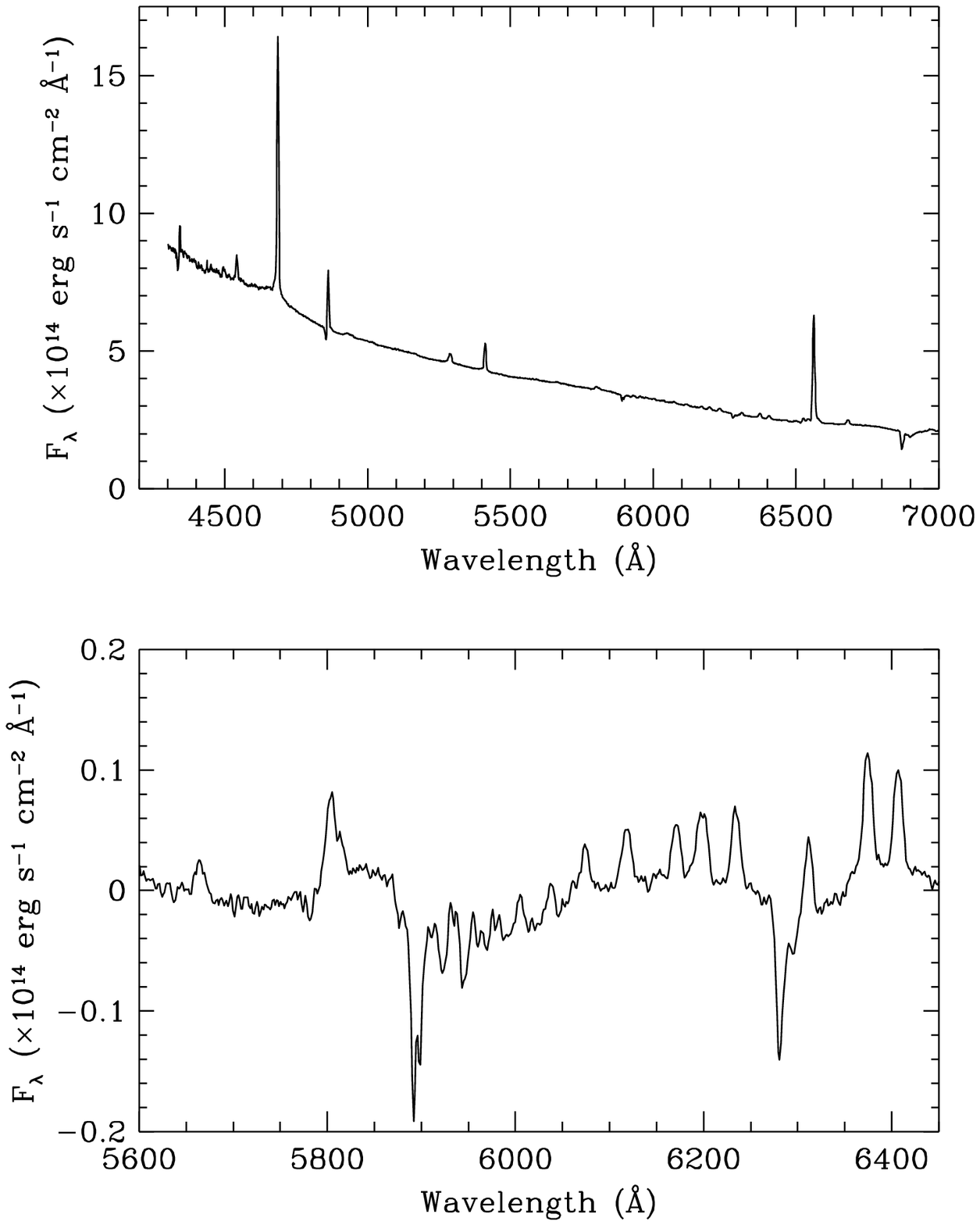]{{\it Top:} Deep spectrum of RXJ0019 in the
emission-line rest frame (see text), prior to continuum removal. {\it
Bottom:} Enlargement of the 5600-6450\AA\ portion of the deep spectrum
after continuum subtraction. Most of the emission lines are the
$n{\rightarrow}5$ transitions of He\,II (see Table 4). \label{fig:0019spec}}

\figcaption[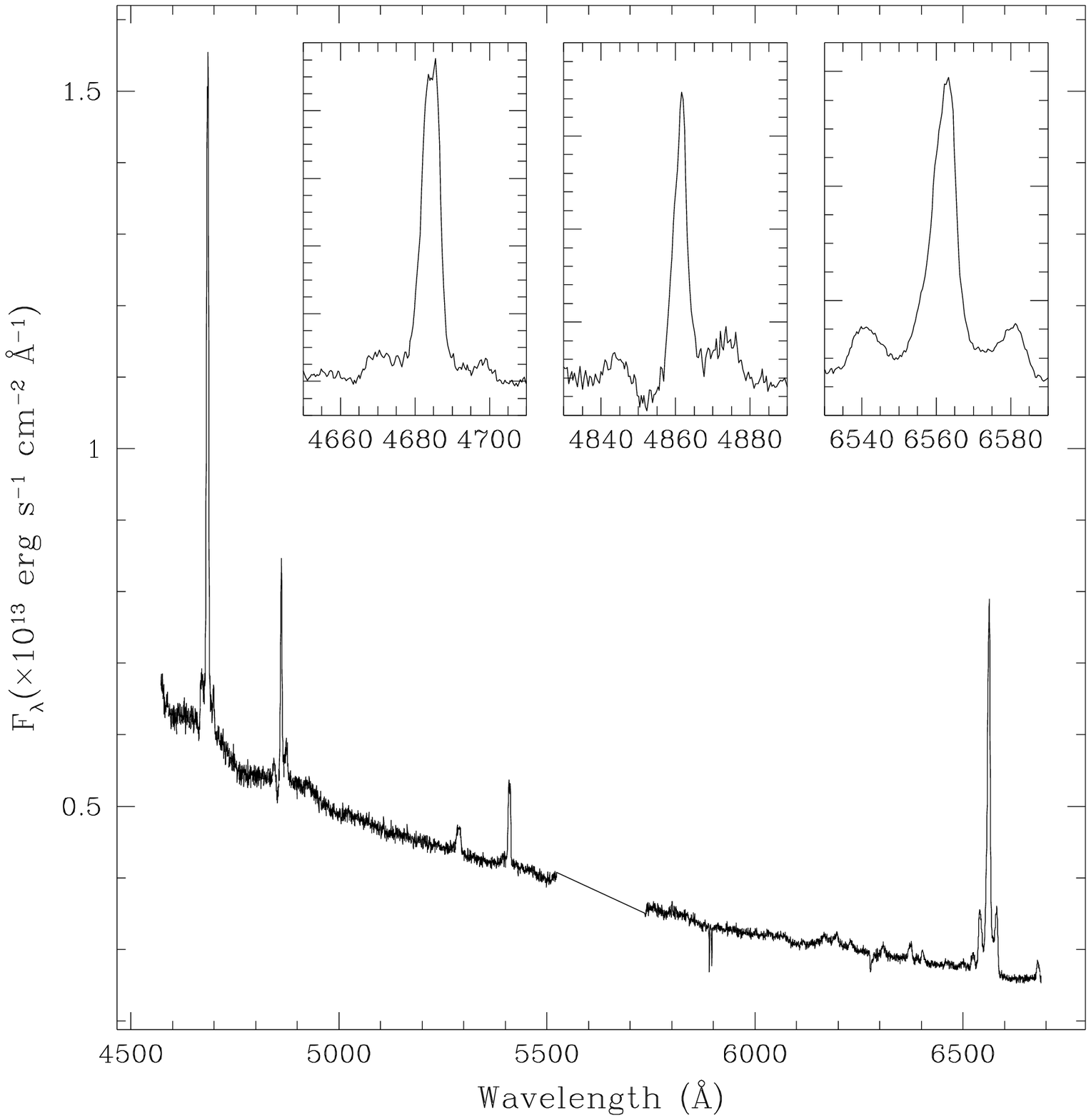]{A fifteen minute observation (per red and blue
exposure) of RXJ0019 made with the FAST spectrograph and the FLWO
1.5$^{\rm m}$ telescope on 12 December 1994.  Signatures of high
velocity lines and P Cygni type absorption are readily seen.  The
insets show enlargements of the He\,II $\lambda$4686, H$\beta$, and
H$\alpha$ profiles (left to right).\label{fig:cfa-outflow}}

\figcaption[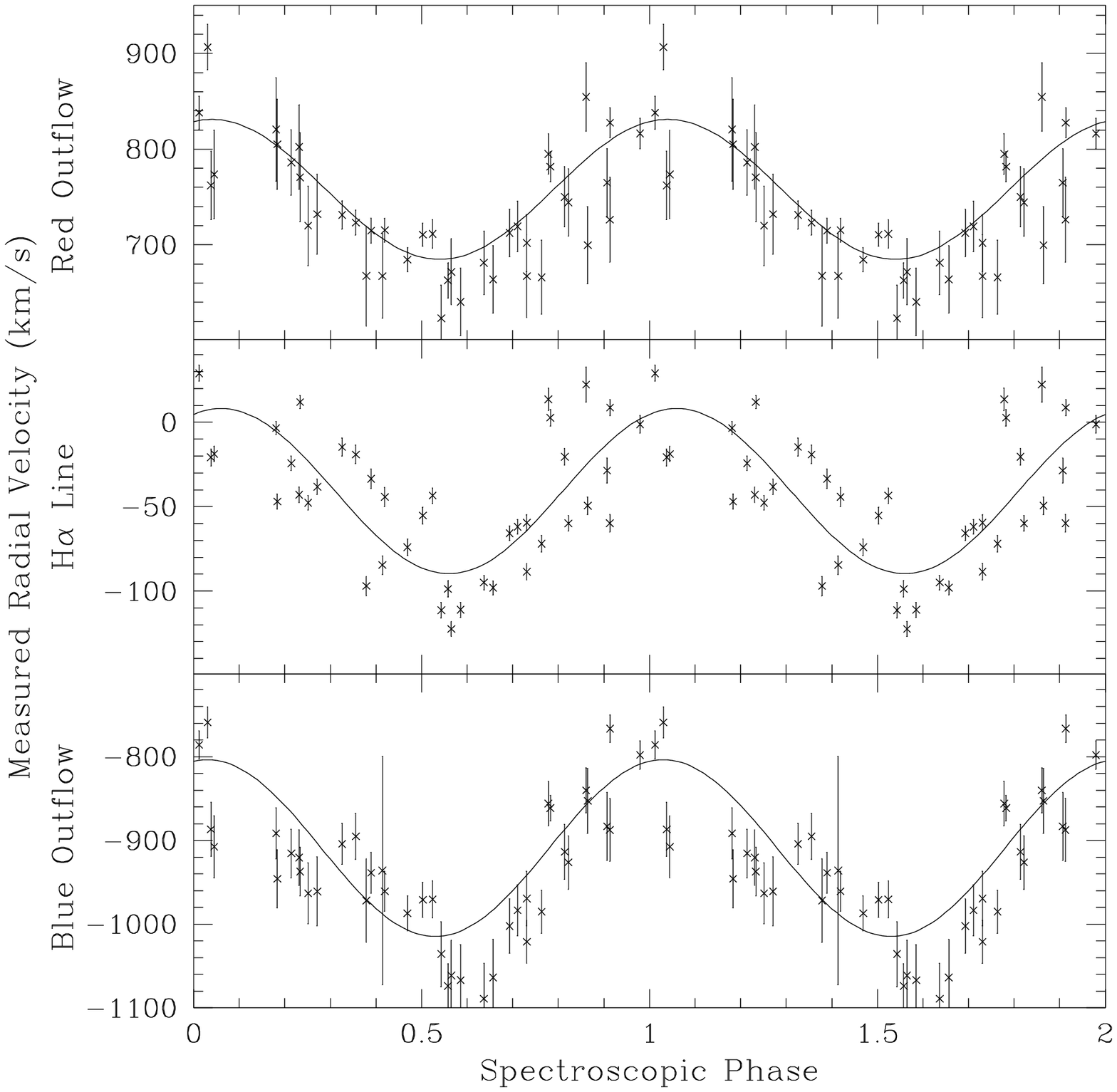]{Velocity of the red outflow {\it (top)},
H$\alpha$ {\it (middle)}, and blue outflow {\it (bottom)} emission as
a function of the spectroscopic phase.  The H$\alpha$ line is used
because it has the strongest outflow components. The solid lines show
the best fit sine curves whose parameters are summarized in Table
5. \label{fig:satphase}}

\figcaption[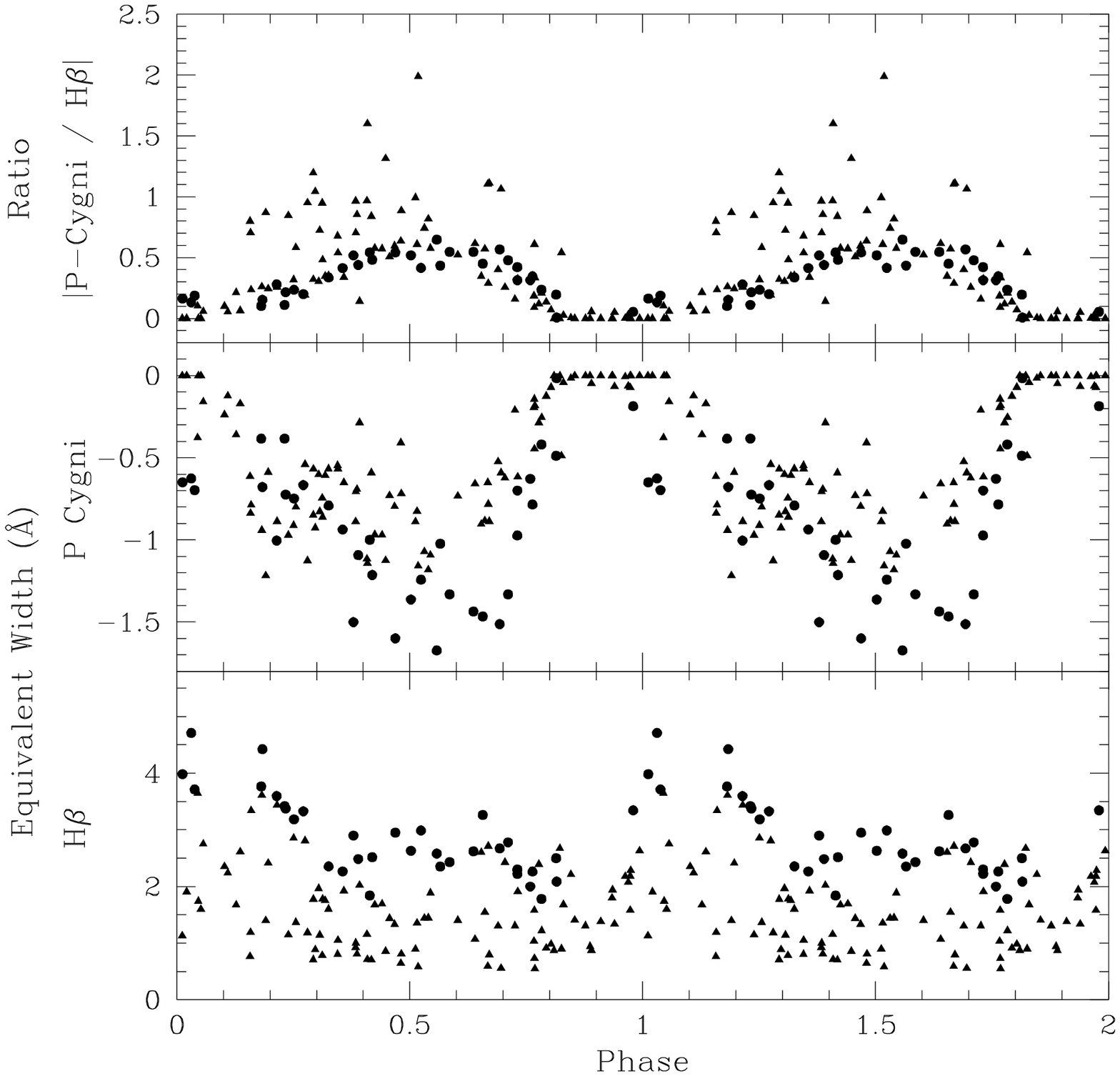]{Equivalent width of the H$\beta$ emission {\it
(bottom)}, P Cygni absorption {\it (middle)}, and the ratio of the two
{\it (top)} as a function of the spectroscopic phase. The dots
(triangles) represent observations with (without) observed high
velocity lines.  H$\beta$ lines are used because the absorption trough
is better defined.\label{fig:pcyg_per}}

\figcaption[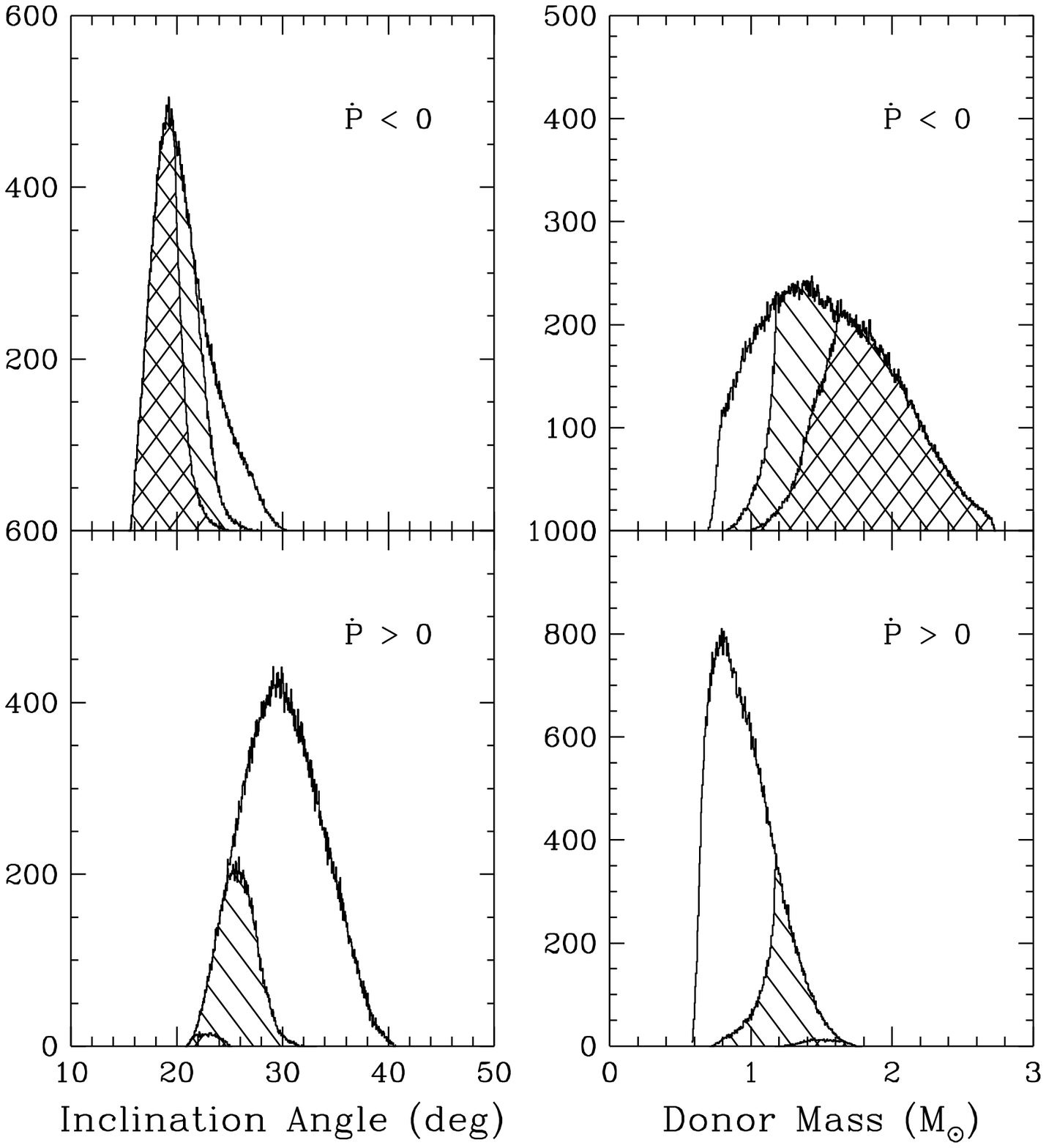]{Histogram of the number of simulated systems as
a function of final orbital inclination (left panels) and final
donor-star mass (right panels). For each case we separate the
systems that reach a 15.85 hr binary period early in their evolution
($\dot{P} < 0$, {\it top}) from those systems that have evolved past
their minimum period ($\dot{P} > 0$, {\it bottom}). The unshaded
histograms represent systems with mass transfer rates below
$3{\times}10^{-8}$ \mspyr, while the two levels of shaded histograms
represent mass transfer rates above $3{\times}10^{-8}$ and $10^{-7}$
\mspyr\, respectively. \label{fig:incline_hist}}

\figcaption[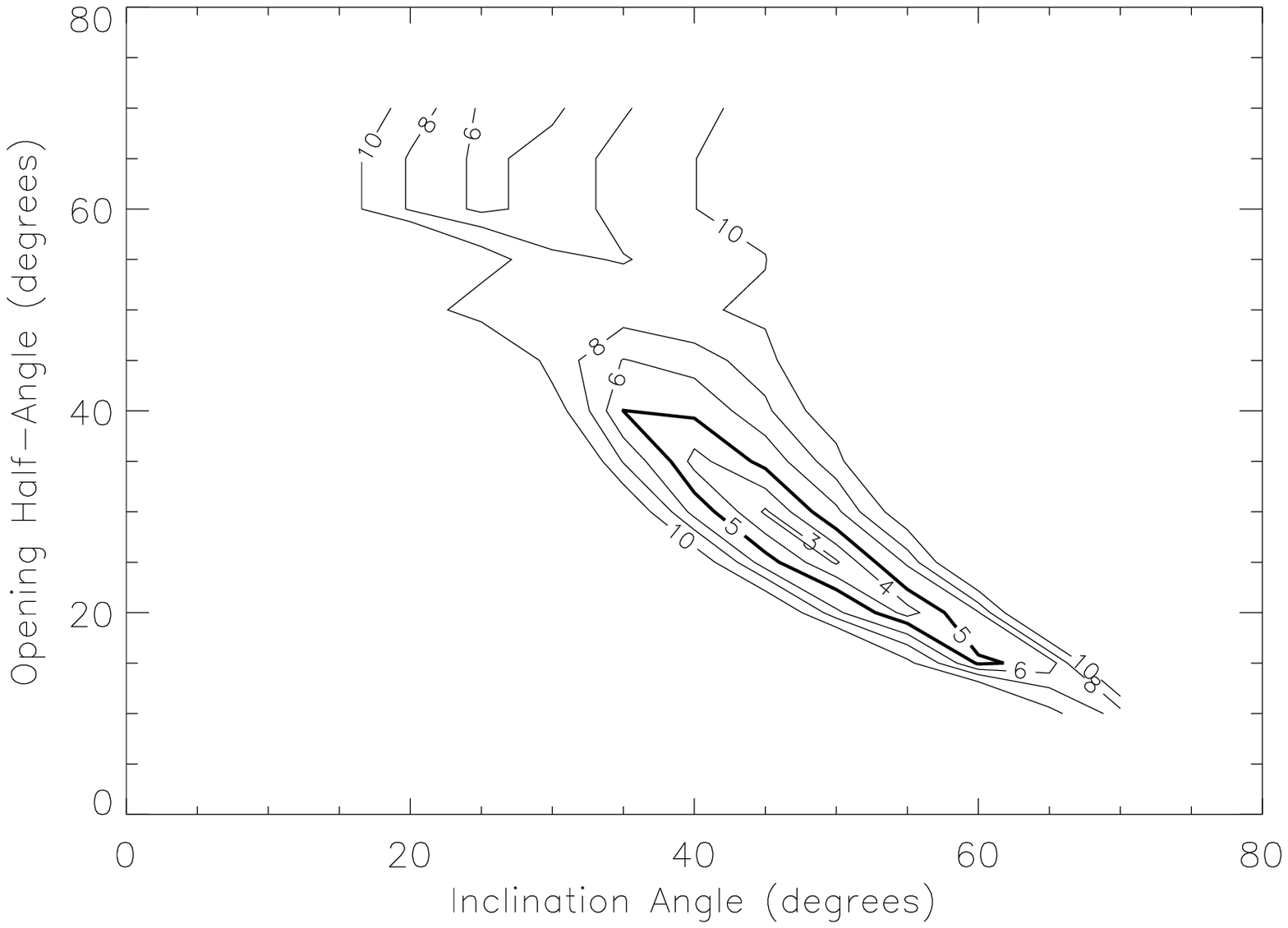]{Contours of the best-fit value of reduced 
chi square as a function of opening half-angle of the jet and the 
inclination angle of the binary system.\label{fig:alpha_incl}}

% ---------------------------------------------------------------
% -------------------     TABLES                 ----------------
% ---------------------------------------------------------------

\clearpage

\begin{deluxetable}{cccccccc}
\footnotesize
\tablecaption{Observations of RXJ0019.8+2156 at MDM \label{tab:0019obs}}
\tablewidth{0pt}
\tablehead{
\colhead{Observing}& \colhead{JD--2400000} & 
\colhead{JD--2400000} & \colhead{Exposure}&
\colhead{Bandpass}&\colhead{Number of}&
\colhead{Number of} & \colhead{Number of} \\
\colhead{Run} & \colhead{Start} & \colhead{End}& \colhead{(sec)} &
 \colhead{(\AA)} & \colhead{Nights} & \colhead{Pointings} & \colhead{Spectra}
}
\startdata
Dec 94  & 49696.693 & 49700.645 & 300 & 3800--6200 &  4 &  4 &  33 \\
Jun 95  & 49875.973 & 49881.932 & 100 & 4100--7200 &  6 &  8 &  88 \\
Oct 95  & 49991.667 & 49996.958 & 200 & 4600--7575 &  6 & 28 & 140 \\
Nov 95  & 50025.694 & 50026.711 & 200 & 3885--6300 &  2 &  4 &  23 \\
Dec 95a & 50050.580 & 50051.802 & 300 & 4500--7425 &  2 &  6 &  28 \\
Dec 95b & 50052.561 & 50055.740 & 300 & 4025--7110 &  4 & 13 &  60 \\
Dec 95c & 50056.646 & 50065.738 & 300 & 4280--7125 & 10 & 18 & 115 \\
Jan 96  & 50086.572 & 50094.566 & 300 & 4280--7140 &  9 & 14 &  87 \\
May 96a & 50236.985 & 50241.967 & 200 & 4521--7485 &  6 &  6 &  17 \\
May 96b & 50249.974 & 50270.954 & 200 & 4536--7485 & 22 & 16 &  91 \\
Dec 96  & 50448.601 & 50454.607 & 300 & 4285--7150 &  6 &  6 &  48 \\
Feb 97  & 50465.630 & 50470.566 & 200 & 4460--7400 &  6 &  6 &  27 \\
\enddata
\end{deluxetable}

\clearpage

\begin{deluxetable}{ccc}
\footnotesize
\tablecaption{Observations of RXJ0019.8+2156 at FLWO
\label{tab:0019obscfa}}
\tablewidth{0pt}
\tablehead{
\colhead{JD - 2400000} & \colhead{Exposure} & \colhead{Bandpass}\\
\colhead{(days)} & \colhead{(sec)} & \colhead{(\AA)}
}
\startdata
49655.738160 & 1200 & 4569--5520 \\
49656.755360 & 900 & 4574--5344 \\
49656.792080 & 900 & 5733--6687 \\
49662.797710 & 900 & 4570--5523 \\
49662.819210 & 900 & 5729--6683 \\
49686.570080 & 900 & 4572--5523 \\
49686.588220 & 900 & 5738--6684 \\
49689.720000 & 900 & 4571--5522 \\
49689.748870 & 900 & 5733--6687 \\
49694.689250 & 900 & 4572--5523 \\
49694.709460 & 900 & 5734--6688 \\
49695.694650 & 900 & 4570--5521 \\
49695.715140 & 900 & 5733--6686 \\
49720.573440 & 900 & 4574--5529 \\
49720.616370 & 900 & 4574--5528 \\
49720.624670 & 300 & 5741--6687 \\
49720.653580 & 900 & 5741--6687 \\
49724.613470 & 900 & 4573--5524 \\
49724.633840 & 900 & 5736--6689 \\
49747.577730 & 900 & 4571--5522 \\
49747.597220 & 900 & 5694--6682 \\
49751.597410 & 900 & 4572--5523 \\
49751.617640 & 900 & 5752--6727 \\
49754.580720 & 900 & 4573--5528 \\
49754.594210 & 900 & 5744--6725 \\
\enddata
\end{deluxetable}

\clearpage

\begin{deluxetable}{ccc}
\footnotesize
\tablecaption{Orbital Parameters for RXJ0019.8+2156
\label{tab:ephtable}}
\tablewidth{0pt}
\tablehead{
\colhead{Parameter}& \colhead{Results} & \colhead{Reference}
}
\startdata

$P$ (He\,II velocities, days) & $0.6604795 \pm\ 0.0000096$ & This work \\
$T_{S}$ (max. velocity, HJD) & $2450002.2713 \pm\ 0.0046$ & This work \\
$\gamma$ (systemic velocity, km \ps) & $-57.5 \pm\ 2.3$ & This work \\
$K$ (velocity semi-amplitude, km \ps) & $71.2 \pm\ 3.6$ & This work \\
Mass function (\msun) & $0.0247 \pm\ 0.0040$ & This work \\
\\
$P$ (photometric, days) & $0.6604721 \pm\ 0.0000072$ & Will \& Barwig 1996 \\
$T_0$ (photometric min., HDJ) & $2450014.273 \pm 0.007$ & Will \& Barwig 1996 \\
\\
$P$ (photographic, days) & $0.6604565 \pm\ 0.0000015$ & Greiner \& Wenzel 1995 \\
$T_0$ (photometric min., HDJ) & $2435799.247 \pm\ 0.033$ & Greiner \& Wenzel 1995 \\
\enddata
\end{deluxetable}

\clearpage

\begin{deluxetable}{ccccl}
\tablecaption{Lines identified in the rest frame spectrum of
RXJ0019.\label{tab:0019lines}} \footnotesize \tablewidth{0pt}
\tablehead{ \colhead{Obs $\lambda$} & \colhead{Species} &\colhead{
Rest $\lambda$ }& \colhead{EW} & \colhead{Notes}}
\startdata 3924.8 & He\,II (15-4) & 3923.5& 0.29 & \\ 3976.9 &
H$\epsilon$ & 3970.1& 0.22 & Almost all emission absorbed in P Cygni\\
4026.3 & He\,II (13-4)& 4025.6& 0.62& \\ 4109.3 & H$\delta$ & 4101.7 &
& Almost all emission absorbed in P Cygni\\ 4199.7 & He\,II (11-4) &
4199.8 & 0.68 & \\ 4343.5 & H$\gamma$ & 4340.5 & 0.43 & Strong P Cygni
Profile\\ 4542.1 & He\,II (9-4) & 4541.6 & 0.94 & \\ 4686.1 & He\,II
(4-3) & 4685.7 & 9.72& \\ 4862.3 & H$\beta$& 4861.3 & 1.61 & Strong P
Cygni Profile\\ 4930.4 & & & 0.33 & \\ 4947.7 & & & 0.15 & \\ 5290.2 &
O\,VI & 5291 & 0.90 & \\ 5412.0 & He\,II (7-4) & 5412.0 & 1.74 & \\
5665.2 & & & 0.14 & \\ 5803.0 & C\,IV & 5801.5 & 0.31 & \\ 5814.2 &
C\,IV & 5812.1 & 0.12 & \\ 6004.5 & He\,II (22-5) & 6007 & 0.11 & \\
6037.2 & He\,II (21-5) & 6036.7 & 0.11 & \\ 6074.3 & He\,II (20-5) &
6074.1 & 0.17 & Part of blend?\\ 6117.8 & He\,II (19-5) & 6118.2 &
0.23 & \\ 6170.6 & He\,II (18-5) & 6170.6 & 0.18 & \\ 6200.7 & & &
0.24 & \\ 6234.3 & He\,II (17-5) & 6233.8 & 0.32 & \\ 6311.7 & He\,II
(16-5) & 6310.8 & 0.30 & At edge of absorption band\\ 6374.5 & Fe\,X &
6374.5 & 0.52& \\ 6406.9 & He\,II (15-5) & 6406.3 & 0.43 & \\ 6502.7 &
& & 0.42 & \\ 6528.1 & He\,II& 6527.9 & 0.60& \\ 6543.9 & & &1.1 & \\
6563.2 & H$\alpha$& 6562.8 & 13.84 & P Cygni absorption present\\
6580.2 & & &1.35 & \\ 6683.3 & He\,II (13-5) & 6683.2& 0.75 & \\
6975.0 & & & 0.73& \\ 7219.8 & & & 3.00 & \\ \enddata
\end{deluxetable}

\clearpage

\begin{deluxetable}{ccccc}
\tablecaption{Fitted parameters to the emission-line velocity
curves.\label{tab:ha_velfits}} 
\tablewidth{0pt} 
\tablehead{
\colhead{Feature}
& \colhead{$\gamma$\tablenotemark{a}} &\colhead{$\gamma - \gamma_{He\,II}$\tablenotemark{a}}
&\colhead{Phase\tablenotemark{a}} & \colhead{$K$\tablenotemark{a}} \\ \colhead{ }  &
\colhead{(km \ps)} & \colhead{(km \ps)} & \colhead{ }& \colhead{(km \ps)} 
}
\startdata
Red & $758.0 \pm\ 10.0$ & $815.5 \pm\ 10.0$ & $0.787 \pm\ 0.032$ & $73.1 \pm\ 14.5$ \\
Central & $-40.7 \pm\ 7.2$ & $16.8 \pm\ 7.2$ & $0.809 \pm\ 0.033$ & $48.9 \pm\ 10.44$ \\
Blue & $-909.2 \pm\ 14.9$ & $-851.9 \pm\ 14.9$ & $0.775 \pm\ 0.032$ & $105.5 \pm\ 20.7$ \\
He\,II $\lambda$4686 & $-57.5 \pm\ 2.3$ & $0.00$ & $0.750 \pm\ 0.007$ &
$71.2 \pm\ 3.6$ \\
\enddata
\tablenotetext{a}{The quoted error bars represent the 1$\sigma$ confidence
limit after scaling the error bars on individual measurements such
that  $\chi^2_\nu = 1$.}

\end{deluxetable}

\clearpage 
\plotone{brrm_1.ps}

\clearpage 
\plotone{brrm_2.ps}

\clearpage 
\plotone{brrm_3.ps}

\clearpage 
\plotone{brrm_4.ps}

\clearpage 
\plotone{brrm_5.ps}

\clearpage 
\plotone{brrm_6.ps}

\clearpage 
\plotone{brrm_7.ps}

\clearpage
\plotone{brrm_8.ps}

\clearpage
\plotone{brrm_9.ps}

\end{document}